\documentclass[prl,twocolumn,preprintnumbers,amsmath,amssymb]{revtex4}

\usepackage{amsmath}
\usepackage{graphicx}
\usepackage{dcolumn}
\usepackage{bm}
\usepackage[dvips]{color}

\begin{document}

\title{Amplification of the photocurrent in SiO$_2$(Co)/GaAs heterostructure induced by magnetic field in the avalanche regime}
\author{V.~V.~Pavlov$^{1}$}
\author{L.~V.~Lutsev$^{1}$}
\author{P.~A.~Usachev$^{1}$}
\author{A.~A.~Astretsov$^{1,2}$}
\author{A.~I.~Stognij$^{3}$}
\author{N.~N.~Novitskii$^{3}$}
\author{R.~V.~Pisarev$^{1}$}
\affiliation{$^{1}$Ioffe Physical-Technical Institute, Russian
Academy of Sciences, 194021 St. Petersburg, Russia}
\affiliation{$^{2}$Academic University –-- Nanotechnology Research
and Education Centre, Russian Academy of Sciences, 194021, St.
Petersburg, Russia}
\affiliation{$^{3}$Scientific and Practical
Materials Research Centre, National Academy of Sciences of Belarus,
220072, Minsk, Belarus}
\date{\today}

\begin{abstract}
Amplification of the photocurrent in heterostructures of silicon
dioxide films containing cobalt nanoparticles grown on gallium
arsenide SiO$_2$(Co)/GaAs has been observed in magnetic field in the
avalanche regime. While the avalanche process is suppressed by the
magnetic field and the current decreases, for photon energy $E$
greater than the GaAs bandgap energy the photocurrent significantly
increases. The amplification reaches 9.5 for $E = 1.50$~eV. The
effect of the photocurrent amplification is explained by the
spin-dependent recombination process at deep impurity centers in
GaAs.
\end{abstract}

 \maketitle

\email{l_lutsev@mail.ru}

The explosive growth of the quantum-information science involving
encoding, communication, manipulation, and measurement of
information using quantum-mechanical objects leads to the intensive
research of single-photon avalanche diodes
(SPADs)~\cite{Hadf09,Eisa11,Fish12}. In addition to the
quantum-information science, SPADs are used in bioluminescence
detection, DNA sequencing, light ranging, picosecond imaging,
single-molecule spectroscopy, diffuse optical tomography,
etc~\cite{Iso95,Rech04,Pell00,Stell04,Mich07,Piff08}. SPADs have
high detection efficiencies, high sensitivities, low dark count
rates, and low jitter times~\cite{Hadf09,Eisa11}. But, with the
above-mentioned advantages, these devices possess the deficiency --
high recovery times, during which detectors are unable to register
photons. High recovery times lead to limitation the ability to
resolve photon number. This problem can be resolved, if incident
photons not only create photo-induced carriers, but influence on the
avalanche feedback during the recovery time. In this case, one can
evaluate number of photons by means of a current change flowing
through an avalanche detector.

The action of incident photons on the avalanche feedback can be
realized in metal-dielectric heterostructures composed of thin film
of amorphous silicon dioxide with cobalt nanoparticles deposited on
$n$-type gallium arsenide substrates SiO$_2$(Co)/GaAs. The avalanche
process in SiO$_2$(Co)/GaAs heterostructures is initiated by
electrons injected from the SiO$_2$(Co) film~\cite{Lut09}. The
avalanche positive feedback is caused by holes generated by the
impact ionization and moved to the potential barrier nearby the
interface~\cite{Lut06a}. Due to the positive feedback, the extremely
large magnetoresistance and the negative photoconductance have been
observed~\cite{Lut09,Lut05,Lut12}.

In order to reach action of light on the avalanche positive feedback
and, therefore, to get high changes of the photocurrent during the
recovery process, which is equal to nanoseconds, the lifetime of
photo-induced holes acted on the avalanche feedback must equal or
greater than the recovery time. In this connection, it is need to
note that the recombination of the charge carriers in GaAs is
related mainly to nonradiative electron transitions through the
localized electron levels in the forbidden energy
band~\cite{Mill80,Pag84,Bar96,Kal2005,Grange07,Ivch2010}. The rate
of electron-hole recombination is determined by the spin-dependent
recombination (SDR) through the localized electron levels in the
semiconductor and depends on the spin states of both paramagnetic
centers and free carriers. The recombination rate can be changed by
an applied magnetic field. In a magnetic field electron-hole
recombination is suppressed and additional free electrons and holes
appear in the conduction band and in the valence band, respectively.
As a result, photo-induced holes are accumulated in the barrier and
the avalanche feedback can be manipulated by the light. The effect
is accompanied by an amplification of the photocurrent.

In this paper, we study the expected amplification of the
photocurrent in SiO$_2$(Co)/GaAs heterostructures in the avalanche
regime. The amplification is observed in magnetic field for light
with the photon energy $E$ greater than the energy of the GaAs
bandgap. For studied heterostructures the photocurrent amplification
reaches 9.5 for $E = 1.50$~eV at the applied magnetic field $H=
1.65$~kOe.

Experiments were performed on heterostructures
(SiO$_2$)$_{100-x}$Co$_x$/GaAs [the abridgeâ notation is
SiO$_2$(Co)/GaAs]. $n$-GaAs substrates with the thickness of 0.4\,mm
were of the (001)-orientation type. Electrical resistivity of GaAs
chips was equal to 1.0$\times 10^5$\,$\Omega\cdot$cm. Prior to the
deposition process, substrates were polished by a low-energy oxygen
ion beam~\cite{Stog02}. The SiO$_2$(Co) films with the thickness of
40 nm were prepared by the ion-beam deposition technique using a
composite cobalt-quartz target onto GaAs substrates heated to
200$^{\circ}$C. The Co concentration in SiO$_2$ matrix was specified
by a relation of cobalt and quartz surface areas. The film
composition was determined by the nuclear physical methods of
element analysis based on spectra of the Rutherford backscattering
for deuterons and nuclear reactions with deuterons. The average size
of Co particles was determined by the small-angle X-ray scattering
and increased with the growth of the concentration $x$: from 3.0~nm
at $x$ = 45 at.\% to 4.0~nm at $x$ = 71 at.\%. A protective Au layer
of the thickness of 3-5 nm has been sputtered on SiO$_2$(Co) films.
The Au layer formed one ohmic contact in experiments, while the
second contact was on the GaAs substrate. Since the Co-concentration
is in the range 45 - 71\,at.\% and enables the percolation
threshold, therefore, the film resistivity of 2.0-1.0$\times
10^2$\,$\Omega\cdot$cm is much smaller than the resistivity of the
GaAs. In this case, the applied voltage $U$ primarily falls on the
GaAs substrates.

\begin{figure}
\begin{center}
\includegraphics*[scale=0.42]{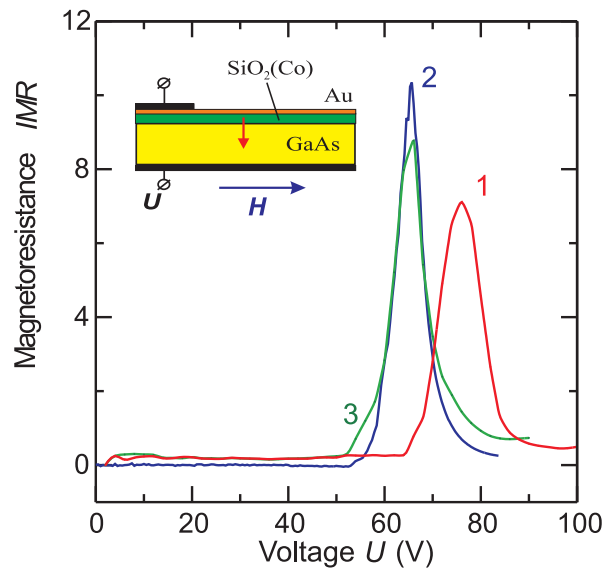}
\end{center}
\caption{(Color online) Injection magnetoresistance ratio {\it IMR}
as a function of the voltage $U$ in the heterostructure
SiO$_2$(Co)/GaAs with Co concentration (1) 45 at.\%, (2) 60 at.\%
and (3) 71 at.\% at magnetic field of 2.1~kOe. Inset shows a scheme
of experimental geometry. } \label{Fig1}
\end{figure}

\begin{figure}
\begin{center}
\includegraphics*[scale=0.42]{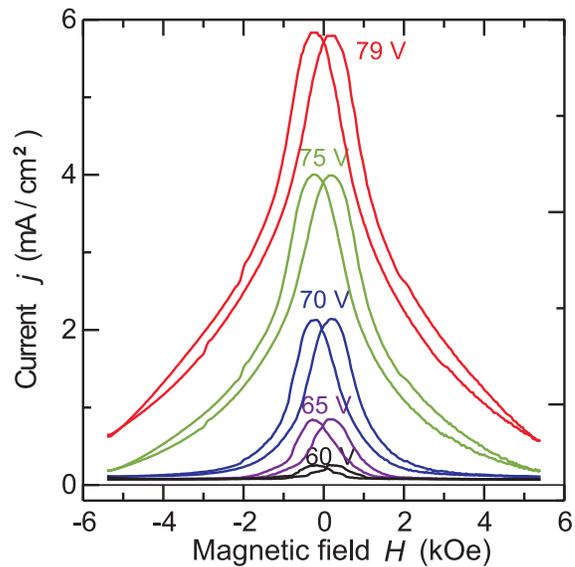}
\end{center}
\caption{(Color online) Current $j$ flowing in the SiO${
}_2$(Co)/GaAs structure with 60~at.\% Co versus the magnetic field
$H$ at room temperature at different voltages $U$. } \label{Fig2}
\end{figure}

Since, the studying action of light on the avalanche is caused by
the SDR and depends on a magnetic field, we choose samples with the
highest influence of magnetic field on the current flowing in
SiO$_2$(Co)/GaAs heterostructures in the avalanche regime and,
therefore, with the highest magnetoresistance coefficient.
Magnetoresistance and influence of the magnetic field on the current
at room temperature are presented in Fig. \ref{Fig1} and Fig.
\ref{Fig2}. Electrons are injected from the granular film into the
GaAs. Figure \ref{Fig1} illustrates the magnetoresistance effect as
a function of applied voltage. The injection magnetoresistance
coefficient {\it IMR} is defined as the ratio~\cite{Lut09,Lut05}

\[{\mbox{\it IMR}}= \frac{R(H)-R(0)}{R(0)}= \frac{j(0)-j(H)}{j(H)},\]

\noindent where $R(0)$ and $R(H)$ are the resistances of the SiO${
}_2$(Co)/GaAs heterostructure without a field and in the magnetic
field $H$, respectively; $j(0)$ and $j(H)$ are the current densities
flowing in the heterostructure in the absence of a magnetic field
and in the field $H$. The magnetic field $H$ is equal to 2.1~kOe and
is parallel to the film surface. For $U >$ 52~V, a sharp increase in
the current due to the process of impact ionization is observed. The
applied magnetic field suppresses the avalanche process and the
current flowing in the SiO$_2$(Co)/GaAs heterostructure decreases
(Fig. \ref{Fig2}). The suppression of the avalanche process causes
to the sharp growth of the {\it IMR} coefficient. As we can see from
Fig. \ref{Fig2} the influence of the magnetic field on the current
is characterized by the hysteresis loop structure.

\begin{figure}
\begin{center}
\includegraphics*[scale=0.75]{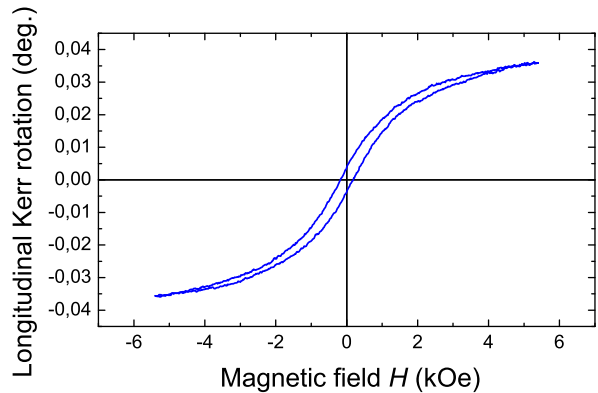}
\end{center}
\caption{(Color online) Hysteresis loops of the SiO${ }_2$(Co)/GaAs
heterostructure with 60~at.\% Co measured by the longitudinal MO
Kerr effect. } \label{Fig3}
\end{figure}

Magnetic characterization of SiO${ }_2$(Co)/GaAs heterostructures in
magnetic field was done by means of the longitudinal magneto-optical
Kerr effect (MOKE). Figure \ref{Fig3} shows magnetic hysteresis
loops for the SiO${ }_2$(Co)/GaAs structure with 60~at.\% Co
measured by MOKE (a magnetic field is aligned along the sample
surface and in the plane of incidence) using a He-Ne laser with a
wavelength of 632.8~nm. The value of incidence angles was of 45~deg.
It is need to note that the SiO${ }_2$(Co) film is in the
ferromagnetic state. Taking into account the magnetoresistance
values and magnetic characteristics of heterostructures presented in
Fig. \ref{Fig1}-\ref{Fig3}, we chose the SiO${ }_2$(Co)/GaAs
heterostructure with 60~at.\% Co with the highest magnetoresistance
ratio {\it IMR} for study the influence of the magnetic field on the
photocurrent.

\begin{figure}
\begin{center}
\includegraphics*[scale=0.57]{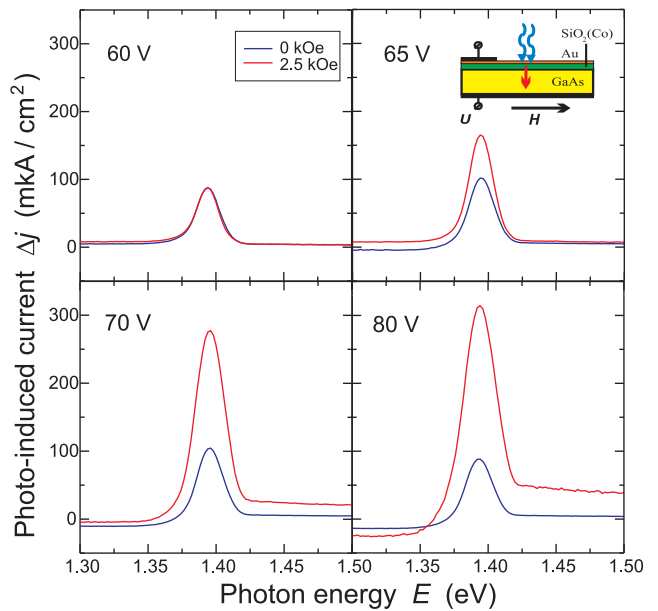}
\end{center}
\caption{(Color online) Spectral dependencies of the photo-induced
current $\Delta j$ in SiO$_2$(Co)/GaAs heterostructure with 60~at.\%
Co caused by the linear-polarized light radiation at different
voltages $U$ without a magnetic field and in the field $H=$ 2.5~kOe.
} \label{Fig4}
\end{figure}

\begin{figure}
\begin{center}
\includegraphics*[scale=0.75]{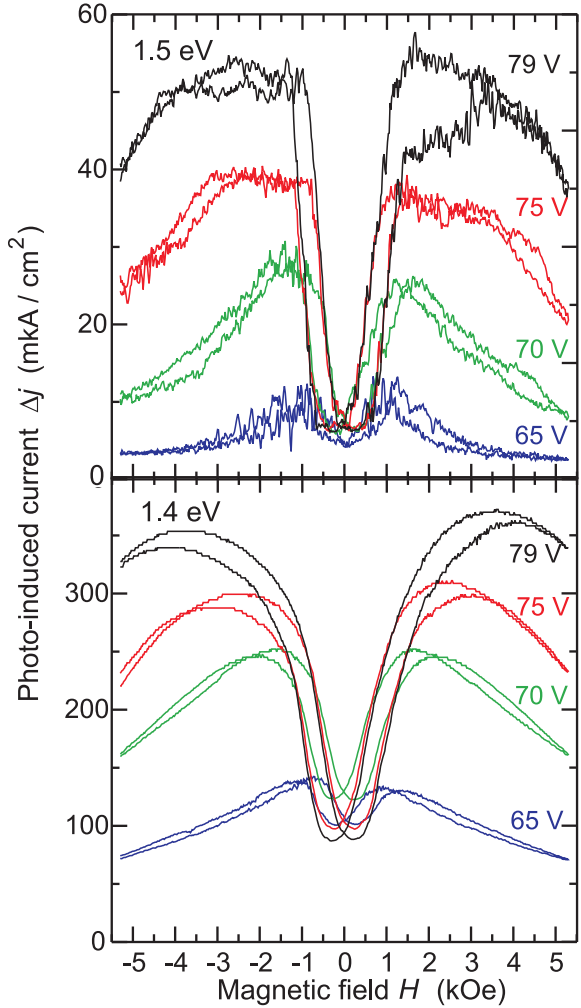}
\end{center}
\caption{(Color online) Photo-induced current $\Delta j$ in
SiO$_2$(Co)/GaAs heterostructure with 60~at.\% Co caused by the
linear-polarized light with photon energies $E =$ 1.5\,eV and
1.4\,eV versus the magnetic field $H$ at different voltages $U$ at
room temperature. } \label{Fig5}
\end{figure}

In optical experiments for measurements of the photo-induced changes
of electrical current (the abridged notation is photo-induced
current) $\Delta j$ in the SiO$_2$(Co)/GaAs heterostructure, we used
the lock-in technique with an intensity modulation of the light. The
light beam was modulated by a chopper working at the frequency of
40~Hz and was exposed on the Au contact layer. Light penetrates
through the SiO$_2$(Co) film and reaches the GaAs substrate. The
light intensity is equal to 1.0~mW/cm$^2$. Spectral dependencies of
the photo-induced current $\Delta j$ caused by the linear-polarized
light at different voltages $U$ without a magnetic field and in the
field $H=$ 2.5~kOe are presented in Figure \ref{Fig4}. At photon
energies $E>$ 1.36\,eV and at voltages $U\geq$ 65~V, in the presence
of the magnetic field the photocurrent increases. If the photon
energy $E <$ 1.36\,eV, at high voltages the negative
photoconductance is observed~\cite{Lut12}. Figure \ref{Fig5} shows
the photo-induced current $\Delta j$ caused by the light radiation
with photon energies $E =$ 1.50\,eV and 1.40\,eV versus the magnetic
field $H$ at different applied voltages at room temperature. In the
range [-1, 1~kOe] the current $\Delta j$ is closed to the square-law
dependence on the field $H$. The photo-induced current reaches
higher values at high voltages $U$ applied to the SiO$_2$(Co)/GaAs
heterostructure. For the studied heterostructure SiO$_2$(Co)/GaAs
with 60~at.\% Co the photocurrent amplification is 9.5 for the
photon energy $E = 1.50$~eV at $U=$ 79\,V at the applied magnetic
field $H= 1.65$~kOe. The subsequent magnetic field growth leads to
the avalanche suppression and to a photocurrent decrease.

\begin{figure}
\begin{center}
\includegraphics*[scale=0.47]{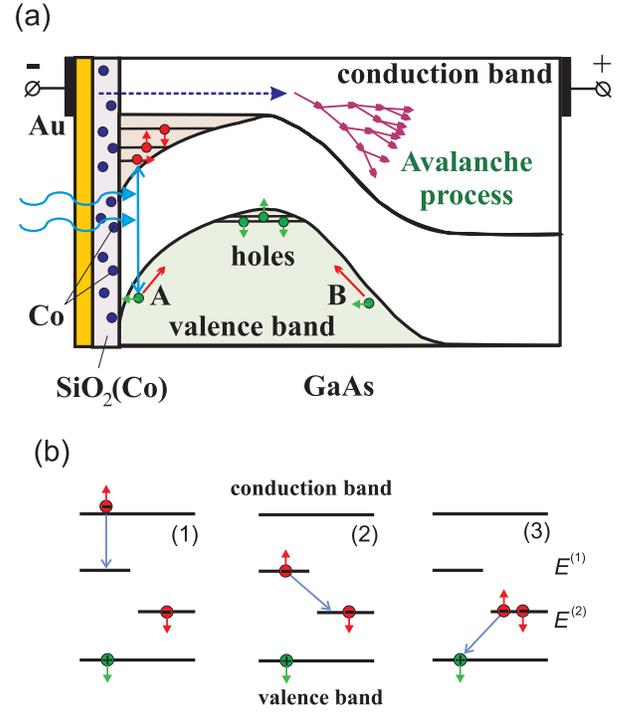}
\end{center}
\caption{ (Color online) (a) Schematic band diagram of the GaAs in
the SiO$_2$(Co)/GaAs heterostructure at the applied electrical field
in the avalanche regime. $A$ marks holes produced by light with the
photon energy greater than the GaAs bandgap energy, $B$ marks holes
produced by the impact ionization. (b) Process
$(1)\rightarrow(2)\rightarrow(3)$ of the recombination of charge
carriers through a pair of localized electron levels (after Kaplan
et al. \cite{Kapl78} and Barabanov et al. \cite{Bar96}).}
\label{Fig6}
\end{figure}

To explain experimental dependencies, we consider electron-hole
processes in SiO$_2$(Co)/GaAs heterostructures in the avalanche
regime. The schematic band diagram of the GaAs in the
SiO$_2$(Co)/GaAs heterostructure at the applied electrical field in
the avalanche regime is shown in Fig. \ref{Fig6}a. The
semiconductor-film interface forms a potential barrier, as a
consequence, electrons are accumulated between the
semiconductor-film interface and the potential maximum of
barrier~\cite{Lut09,Lut06a}. Without a light radiation, the injected
electrons excite the process of impact ionization producing holes in
the valence band. The holes move toward the barrier and are
accumulated there. This lowers the potential maximum of barrier and
increases the electron current leading to an enhancement of the
avalanche. Due to the formed positive feedback, a small variation of
the potential maximum leads to strong changes in the electron
current.

The interface region of the GaAs contains oxygen ions leaved after
the polished process. According to Refs.\,\cite{Lin76,Yu84} in
addition to the EL2 defect level there are oxygen-ion levels in the
GaAs bandgap with energies $E_1= 0.48$\,eV, $E_2=0.74$\,eV,
$E_3=1.0$\,eV, and $E_4=1.25$\,eV counted from the bottom of the
conduction band. At room temperature levels with energies $E_2$,
$E_3$ and $E_4$ are occupied by electrons. Conductivity in the
conduction band is determined by thermally activated electrons from
the level $E_1=0.48$\,eV. These localized levels are of great
importance for the SDR -- conduction electrons and holes created in
excess recombine through intermediate recombination center.

Let us consider the photocurrent amplification for the light with
the photon energy $E>E_g$, where $E_g$ is the bandgap energy. The
light with the photon energy $E>E_g$ is absorbed in the
semiconductor region near the interface and creates electrons in the
conduction band and holes in the valence band in the interface
region. According to theoretical models of the SDR developed in
\cite{Bar96,Ivch2010,Lep72,White77,Kapl78,Vlas95,Bohme2002}, the
recombination event is preceded by the trapping of conduction
electrons to localized electron levels (Fig. \ref{Fig6}b). In order
to find photocurrent dependencies on the magnetic field, we consider
two electrons on a pair of localized levels with different energies
$E^{(1)}$ and $E^{(2)}$. The spin conservation law permits an
electron transition from the first level $E^{(1)}$ to the second
level $E^{(2)}$ and a further recombination occurs if the electron
pair on the levels $E^{(1)}$ and $E^{(2)}$ is in the single state
$|S\rangle=\frac{1}{\sqrt{2}}(|\uparrow\downarrow\rangle-
|\downarrow\uparrow\rangle)$. By contrast, the transition
$E^{(1)}\rightarrow E^{(2)}$ is prohibited, when the triplet states
$|T_0\rangle=\frac{1}{\sqrt{2}}(|\uparrow\downarrow\rangle+
|\downarrow\uparrow\rangle)$, $|T_{+}\rangle=
|\uparrow\uparrow\rangle$ and $|T_{-}\rangle=
|\downarrow\downarrow\rangle$ of the pair appears. In this case, a
further recombination is highly suppressed. Under the assumption
that triplet states cannot recombine and singlet states recombines
with a probability $r_s$, the average recombination probability of
electrons on levels $E^{(1)}$ and $E^{(2)}$ is written as $R=r_s/4$.

In the presence of the external magnetic field $H$ levels $E^{(1)}$
and $E^{(2)}$ are splitted. Neumann's density operator $\rho$ of
electrons on splitted levels in a thermal equilibrium is represented
in the Boltzmann form~\cite{Bohme2002}

\[\rho_0= \frac{\sum_{i=1}^4\exp(-\varepsilon_i/kT)|i\rangle\langle
i|}{\sum_{i=1}^4\exp(-\varepsilon_i/kT)},\]

\noindent where $T$ is the temperature, $|i\rangle= \{|S\rangle,
|T_0\rangle, |T_{+}\rangle, |T_{-}\rangle\}$, $\varepsilon_s=0$,
$\varepsilon_0=0$, $\varepsilon_{+}= g\mu_BH$, $\varepsilon_{-}=
-g\mu_BH$ are Zeeman energies of singlet and triplet states,
respectively, $g$ is the $g$-factor of electrons on localized
states, $\mu_B$ is the Bohr magneton. We consider a change of the
average recombination probability of electrons on a highly saturated
non-equilibrium state with the density operator

\[\rho_{sat}= \frac14 \sum_{i=1}^4|i\rangle\langle
i|.\]

\noindent Then, the change of the average recombination is

\begin{equation}
\Delta R=r_s{\rm Tr} \left[|S\rangle\langle
S|(\rho_{sat}-\rho_0)\right]= \frac{r_s}{4}
\tanh^2\left(\frac{g\mu_BH}{2kT}\right). \label{eq1}
\end{equation}

\noindent Taking into account that the photoconductivity
$\sigma_{ph}$ and the number of photo-induced electrons $n_e$ in the
conduction band and the number of holes $n_h$ in the valence band
are inversely proportional to the recombination probability $R$ and
$\Delta R\ll R$, in the second order of the Taylor-series expansion
of the relation (\ref{eq1}) we find

\begin{equation}
\frac{\Delta n_e}{n_e}= \frac{\Delta n_h}{n_h}=\frac{\Delta
\sigma_{ph}}{\sigma_{ph}}=\frac{\Delta R}{R}=
\left(\frac{g\mu_BH}{2kT}\right)^2. \label{eq2}
\end{equation}

Additional holes $\Delta n_h$ move toward the barrier and are
accumulated in the barrier region (Fig. \ref{Fig6}a). The barrier
height lowers. This leads to the observed increase of the electron
current and to the enhancement of the avalanche in the applied
magnetic field. Such positive feedback produced by photo-induced
holes and additional photo-induced electrons $\Delta n_e$ result in
strong changes in the electron current $\Delta j$. The
experimentally observed quadratic dependence of the photocurrent
$\Delta j$ from the magnetic field $H$ at low field values (Fig.
\ref{Fig4}) corresponds to the relation (\ref{eq2}) and confirms the
magnetic dependence obtained in theoretical models developed in
\cite{Lep72,White77}. At high values of the magnetic field $H$, the
avalanche is suppressed and the photocurrent decreases.

In summary, we observe significant influence of magnetic field on
photocurrent flowing in SiO$_2$(Co)/GaAs heterostructures in the
avalanche regime. While the avalanche process is suppressed by the
magnetic field and the current decreases, for photon energy $E$
greater than the GaAs bandgap energy the photocurrent significantly
increases. The amplification reaches 9.5 for $E = 1.50$~eV at the
applied magnetic field $H= 1.65$~kOe. The photocurrent amplification
is explained by the spin-dependent recombination through the
localized electron levels in the semiconductor. In a magnetic field
electron-hole recombination is suppressed. This leads to additional
free electrons in the conduction band and holes in the valence band.
As a result, photo-induced holes are accumulated in the barrier, the
photocurrent grows and the avalanche feedback can be manipulated by
the light. The photocurrent amplification effect can be used in
single-photon avalanche diodes to register photons during recovery
times and to resolve photon number.

The authors gratefully acknowledge the assistance of V.M. Lebedev
(PNPI, Gatchina, Leningrad region, Russia) for determination of the
film composition. This work was supported by the Russian Foundation
for Basic Research (Project Nos. 10-02-01008, 10-02-00516,
10-02-90023).

{e-mail: l\_lutsev@mail.ru}

\end{document}